# Soliton-Generating $\tau$-Functions Revisited

Yair Zarmi  
Jacob Blaustein Institutes for Desert Research  
Ben-Gurion University of the Negev  
Midreshet Ben-Gurion, 8499000, Israel

Abstract

Within the framework of the Inverse-Scattering formalism and the Hirota algorithm, soliton solutions of evolution equations are images of $\tau$-functions. Typically, the latter are expressed in terms of exponentials, the arguments of which are linear in the coordinates. Consequently, often, $\tau$-functions are unbounded in space and time. However, they are not unique. Exploitation of their non-uniqueness uncovers physically interesting possibilities:

1) One can construct equivalent $\tau$-functions, which generate the same traditional (Inverse-Scattering/Hirota)) soliton solutions, yet allow for the extension of the family of soliton solutions to a wider, parametric family, in which the traditional solutions are a subset. The parameters are shifts in individual soliton trajectories.

2) When two wave numbers in a multi-soliton solution are made to coincide, the reduction of the solution to one with a lower number of solitons is qualitatively different for solutions that are within the traditional subset and those that are outside this subset.

3) One can construct $\tau$-functions that are bounded in space and time, in terms of which soliton solutions become images of localized sources.

# 1. Introduction

Many well known nonlinear evolution equations provide approximate descriptions of phenomena in physical systems. For example, the KdV equation describes the propagation of waves in (1+1) dimensions on the surface of a shallow water layer [1,2], along a Fermi-Pasta-Ulam chain [3], and of ion acoustic waves in Plasma Physics [4, 5]; the Kadomtsev-Petviashvili II (KP II) equation describes the propagation of waves in (1+2) dimension on the surface of a shallow water layer [6]. Hence, it is important to find the widest possible families of solutions of such equations.

In the Inverse-Scattering/Hirota approach [7-18], soliton solutions of evolution equations are transforms of $\tau$-functions. Typically, the latter are expressed in terms of exponentials, the arguments of which are linear in the coordinates. As such, often, these $\tau$-functions are unbounded in space and time. However, the $\tau$-functions are not unique.

In this paper, this non-uniqueness is exploited in the construction of $\tau$-functions that are equivalent to the traditional (Inverse-Scattering/Hirota) ones. Namely, they generate the same traditional soliton solutions. Studying the structure of these new $\tau$-functions it is found that they actually generate a wider family of soliton solutions, of which the traditional solutions are a subset. In this wider family, the constant shifts in soliton trajectories are the parameters that characterize the solutions. The traditional subset corresponds to shifts with a specific dependence on soliton wave numbers.

The parameters affect the reduction of a multi-soliton solution to one with a lower number of solitons in the limit when two wave numbers are made to coincide. The result is qualitatively different for solutions that are within the traditional subset and those that are outside this subset. Furthermore, one can construct $\tau$-functions that are bounded in space and time. The soliton-solutions then become images of localized sources. For example, in the cases of both the KdV and the KP II equations, a single-soliton solution is the image of a single soliton, and multi-soliton solutions are images of spatially localized entities. Finally, soliton solutions may be viewed as nonlinear mappings of finite discrete lattices.

The case of the KdV equation is discussed in detail in Appendix I and in Section 2. Examples are provided in the cases of the KP II and modified KdV equation in Sections 3 and 4, respectively.

Throughout the paper, any quantity associated with the Inverse-Scattering/Hirota formalism will be called a "traditional" quantity.

**2. The KdV equation**
The soliton solutions of the KdV equation,

$$u_t = 6uu_x + u_{xxx} , \qquad (1)$$

are constructed in terms of a $\tau$-function through [10, 14]

$$u(t,x) = 2\partial_x^2 \log \tau(t,x) . \qquad (2)$$

$\tau(t,x)$ is not unique. Multiplying it by any term of the form $e^{\mu x + f(t)}$ yields an equivalent $\tau$-function that generates the same solutions of the KdV equation.

**2.1 $N$-soliton solution**
In this Section, the $N$-soliton solution is discussed. Examples of 1-, 2-, 3- and 4-soliton solutions are discussed in detail in Sections 2.2 - 2.5. The traditional $\tau$-function is given by [10,14]:

$$\tau_T(t,x) = 1 + \sum_{n=1}^{N} \sum_{1 \le i_1 < i_2 < \cdots < i_n \le N} \left( \prod_{j=1}^{n} \prod_{l=j+1}^{n} \left( \frac{k_{i_l} - k_{i_j}}{k_{i_l} + k_{i_j}} \right)^2 \right) e^{2\sum_{m=1}^{n} \theta_{i_m}} . \qquad (3)$$

$$k_1 < k_2 < \cdots < k_N$$

$$\theta_i = k_i x + \omega_i t + \delta_i , \qquad (4)$$

$$\omega_i = 4k_i^3 . \qquad (5)$$

Here and in the following the subscript $T$ denotes the value in the traditional (Inverse-Scattering/Hirota) formulation. Finally, note that the sum in Eq. (3) contains $2^N$ terms.

### 2.1.1 Equivalent $\tau$-function

In the traditional family of solutions, the constant shifts, $\delta_i$, play a minor role. They determine the shifts in individual soliton trajectories in the *x-t* plane. In the wider family, presented in the following, they become parameters, which, in addition to determining the shifts in soliton trajectories, affect the quantitative and qualitative features of multi-soliton solutions.

The $\tau$-function is not unique. A $\tau$-function, which generates the same *N*-soliton solution as $\tau_T$, is:

$$\tau_E(t,x) = \sum_{\vec{\sigma}} \prod_{i=1}^{N} \prod_{j=i+1}^{N} (k_j - \sigma_i \sigma_j k_i) \cosh \tilde{\theta}_{\vec{\sigma}}$$

$$\tilde{\theta}_{\vec{\sigma}} = \sum_{i=1}^{N} \sigma_i \tilde{\theta}_i \quad , \quad \tilde{\theta}_i = k_i x + \omega_i t + \Delta_i(\vec{k}) \tag{6}$$

$$\left( \vec{k} = \{k_1, k_2, \cdots, k_N\} \quad , \quad \vec{\sigma} = \{\sigma_1, \sigma_2, \cdots, \sigma_N\} \quad , \quad \sigma_1 = +1, \sigma_{i>1} = \pm 1 \right)$$

The subscript *E* indicates that this $\tau$-function is equivalent to the traditional $\tau$-function, $\tau_T$ of Eq. (3). The details of the transformation from $\tau_T$ to $\tau_E$ are presented in Appendix I. Note that $\delta_i$, the constant shifts in the traditional solution, are replaced by wave number dependent shifts, $\Delta_i(\vec{k})$. The values of $\Delta_i(\vec{k})$ in the traditional case are determined by the transformation from $\tau_T$ to $\tau_E$. Examples are presented in Section 2.3,.24.,2.5, and in Appendix I. Finally, note that the sum in Eq. (6) contains $2^{N-1}$ terms.

### 2.1.2 Soliton solutions as images of localized sources
Define

$$\tau_B(t,x) = \frac{1}{\tau_E(t,x)} . \tag{7}$$

$\tau_B$ is bounded throughout the *x-t* plane (hence, the subscript *B*). Furthermore, the transformation

$$u(t,x) = -2 \partial_x^2 \log \tau_B(t,x) \tag{8}$$

generates the *same N*-soliton solution of the KdV equation. Hence, under Eq. (8), the solution is an image of a spatially localized source. In the case of the single-soliton solution, the source itself is a soliton (Section 2.2). In the cases of multi-soliton solutions, it is a hump that is concentrated in the vicinity of the soliton interaction region (see Sections 2.3, 2.4 and 2.5).

### 2.1.3 Parametric family of soliton solutions

The detailed analysis of the two- three- and four- soliton solutions in Sections 2.3, 2.4 and 2.5, respectively, and the results of Appendix I suggest a conjecture, to be presented later on in this Section, regarding properties of the $N$-soliton solution. The case of the five-soliton solution has been also studied, and leads to the same conclusions. It is omitted because its analysis is excessively long and does not add any new information.

In the traditional case, the wave number dependence of the shifts, $\Delta_i(\vec{k})$, is determined by the transformation from $\tau_T$ of Eq. (3) to $\tau_E$ of Eq. (6). This is shown explicitly in the cases of the two- three- four- and $N$-soliton solutions in Sections 2.3, 2.4 and 2.5, and in Appendix I, respectively.

However, in Sections 2.3, 2.4 and 2.5 it is also shown that the equivalent $\tau$-function, $\tau_E$, generates, respectively, two-, three- and four-soliton solutions for *any* values of $\Delta_i(\vec{k})$, not just the values required in the traditional case. This finding has been also confirmed in the case of the five-soliton solution, the details of the computation of which are not reported here. Hence:

Conjecture
*The equivalent $\tau$-function, $\tau_E$ of Eq. (6), generates an N-soliton solution of the KdV equation for any values of the constant shifts, $\Delta_i(\vec{k})$.*

This conjecture implies that $\tau_E$, generates a family of $N$-soliton solutions of the KdV equation, which depend on the $N$ shifts, $\Delta_i(\vec{k})$, as parameters. The traditional solutions are a subset in this family. They are obtained when the shifts, $\Delta_i(\vec{k})$, obtain their traditional values.

Finally, in the traditional case, the wave-number dependent multiplicative factors in $\tau_T$ of Eq. (3) are necessary for ensuring that $\tau_T$ generates an *N*-soliton solution. In the transformation to $\tau_E$ of Eq. (6), these factors determine both the wave number dependent part of the shifts, $\Delta_i(\vec{k})$, as well as the wave-number dependent factors multiplying the hyperbolic cosine functions. $\tau_E$ is equivalent to $\tau_T$. They both generate the same traditional soliton solutions. However, once $\tau_E$ has bee constructed, it is found that the wave-number dependent factors multiplicative factors in $\tau_E$ ensure that it generates soliton solutions of the KdV equation for *any* values of the shifts, $\Delta_i$, be they wave number dependent, as required in the traditional case, or not.

### 2.1.4 Peculiar properties of solutions in *N*-parameter family

It is well know that when two wave numbers coincide, say, $k_2 \to k_1$, in the traditional case, an *N*-soliton solution is reduced to an (*N* − 1)-soliton solution. This can be readily seen from inspection of $\tau_T$ of Eq. (3). In the limit, the number of terms in the sum in Eq. (3) is reduced from $2^N$, corresponding to *N* solitons, to $2^{N-1}$, corresponding to (*N* − 1) solitons. In the equivalent $\tau$-function, $\tau_E$ of Eq. (6), this result is not as obvious. In the traditional case, the shifts, $\Delta_i(\vec{k})$, have a singular dependence on the wave numbers (see Sections 2.3, 2.4 and 2.5 and Appendix I). This singular behavior is responsible to the reduction of an *N*-soliton solution to an (*N* − 1)-soliton one. While this can be seen directly from inspecting Eq. (6) with the aid of Appendix I, it may be easier to see this through the rigorous analysis of the examples of the two-, three- and four-soliton cases.

However, if $\Delta_i(\vec{k})$ lack the singular wave number dependence required in the traditional case, then, when $k_2 \to k_1$, the number of terms in $\tau_E$ of Eq. (6) is reduced from $2^{N-1}$ to $2^{N-3}$. The solution is reduced to an (*N* − 2)-soliton solution! In the following Sections, it is shown explicitly that, when the $\Delta_i(\vec{k})$ are independent of the wave numbers, the limits of the two-, three- and four-soliton solutions are, respectively, zero, one- and two-soliton solutions.

In the case of $N \geq 4$ solitons, there are other possibilities. If some $\Delta_i(\vec{k})$ are constant, while others have the singular structure required in the traditional case, then, when some wave number pairs coincide, the solution may be reduced to $(N - 2)$ solitons, while when other wave number pairs coincide, the solution may be reduced to $(N - 1)$ solitons.

### 2.1.5 Solitons as images of lattices

When soliton solutions are constructed from $\tau_E$ through Eq. (2), or from $\tau_B$ through (8), they can be viewed as nonlinear mappings of a lattice comprised of the $2^{N-1}$ points $\vec{\sigma}$ defined in Eq. (6).

### 2.2 Single-soliton solution

This trivial case is discussed only so as to show the emerging pattern. The traditional $\tau$-function,

$$\tau_T(t,x) = 1 + e^{2(kx + 4k^3t + \delta)} , \qquad (9)$$

where $\delta$ is a constant shift, generates the single-soliton solution:

$$u(t,x) = \frac{2k^2}{\left(\cosh(kx + 4k^3t + \delta)\right)^2} . \qquad (10)$$

Multiplying the expression in Eq. (9) by

$$e^{-(kx + 4k^3t + \delta)/2}$$

yields an equivalent $\tau$-function, which generates the same single-soliton solution through Eq. (2):

$$\tau_E(t,x) = \cosh(kx + 4k^3t + \delta) . \qquad (11)$$

Now choose to replace $\tau_E$ by

$$\tau_B(t,x) = \frac{1}{\tau_E(t,x)} = \frac{1}{\cosh(kx + 4k^3t + \delta)} . \qquad (12)$$

The single soliton solution is then generated by Eq. (8).

Unlike $\tau_T$ and $\tau_E$, $\tau_B$ is bounded in the $x$-$t$ plane. Hence, under Eq. (8), the single-soliton solution of the KdV equation is the image of a spatially bounded entity, the single soliton of Eq. (12).

## 2.2 Two-soliton solution

The traditional $\tau$-function corresponding to a two-soliton solution (wave numbers $k_2 > k_1 > 0$) is

$$\tau_T(t,x) = 1 + e^{2\theta_1} + e^{2\theta_2} + \left(\frac{k_2 - k_1}{k_1 + k_2}\right)^2 e^{2\theta_1} e^{2\theta_2} , \qquad (13)$$

### 2.2.1 Single-parameter family of solutions

Replace Eq. (13) by adding a multiplicative factor:

$$\tau_T' = \left((k_1 + k_2)^2 e^{-\theta_1} e^{-\theta_2}\right) \tau_S . \qquad (14)$$

$\tau_T'$ generates the same soliton solutions through Eq. (2). The result may be expressed in terms of

$$\theta_{++} = \theta_1 + \theta_2 \quad , \quad \theta_{+-} = \theta_1 - \theta_2 \qquad (15)$$

as:

$$\tau_T' = 2(k_1 + k_2)\{(k_2 - k_1)\cosh(\theta_{++} + \alpha) + (k_1 + k_2)\cosh\theta_{+-}\} . \qquad (16)$$

In Eq. (16),

$$\sinh\alpha = \frac{-2 k_1 k_2}{(k_2 - k_1)(k_1 + k_2)} . \qquad (17)$$

As the constant multiplicative factor, $(2(k_1 + k_2))$, on the r.h.s. of Eq. (16) does not affect the definition of $u(t,x)$ in Eq. (2), it can be omitted, resulting in the following equivalent $\tau$-function:

$$\tau_E(t,x) = (k_2 - k_1)\cosh(\theta_{++} + \alpha) + (k_1 + k_2)\cosh\theta_{+-} . \qquad (18)$$

Using $\tau_E$, with Eq. (17) for $\alpha$, in Eq. (2), yields the same two-soliton solution as $\tau_T$ of Eq. (13). However, provided $\omega_i$ are given by Eq. (5), $u(t,x)$ turns out to be a two-soliton solution of the KdV equation for *any* value of $\alpha$, not just that of Eq. (17). Thus, the existence of a single-parameter family of two-soliton solutions has been established, of which the traditional solution is just one member. Finally, the wave-number dependent multiplicative coefficients in Eq. (18) follow the general rule of Eq. (6): The sign within each coefficient is determined by the product $(\sigma_i \sigma_j)$:

$$\begin{aligned}\theta_{++} &: \sigma_1 = +1, \sigma_2 = +1 \rightarrow (k_2 - k_1) = (k_2 - \sigma_1 \sigma_2 k_1) \\ \theta_{+-} &: \sigma_1 = +1, \sigma_2 = -1 \rightarrow (k_2 + k_1) = (k_2 - \sigma_1 \sigma_2 k_1)\end{aligned} . \qquad (19)$$

### 2.2.2 Solution parameters as soliton trajectory shifts

In $\tau_E$ of Eq. (18) the effect of $\alpha$ can be translated into individual shifts of soliton trajectories:

$$\tau_E(t,x) = (k_2 - k_1)\cosh\tilde{\theta}_{++} + (k_1 + k_2)\cosh\tilde{\theta}_{+-}$$

$$\tilde{\theta}_{++} = \tilde{\theta}_1 + \tilde{\theta}_2 \quad , \quad \tilde{\theta}_{+-} = \tilde{\theta}_1 - \tilde{\theta}_2 \tag{20}$$

$$\tilde{\theta}_i = \theta_i + \frac{\alpha}{2} = k_i x + \omega_i t + \Delta_i(\vec{k}) \quad , \quad \left(\Delta_i(\vec{k}) = \delta_i + \frac{\alpha}{2}\right)$$

### 2.2.3 Localized source

Following Eq. (7) and defining a new $\tau$-function:

$$\tau_B(t,x) = \frac{1}{\tau_E(t,x)} = \frac{1}{(k_2 - k_1)\cosh\tilde{\theta}_{++} + (k_1 + k_2)\cosh\tilde{\theta}_{+-}} \,, \tag{21}$$

Eq. (8) generates the same single-parameter family of two-soliton solutions. However, unlike $\tau_T$ and $\tau_E$, $\tau_B$ is localized in the $x$-$t$ plane. Hence, under Eq. (8), the two-soliton solution is the image of a localized source. The peak of the source is at $\tilde{\theta}_{++} = \tilde{\theta}_{+-} = 0$, and its widths in the $\tilde{\theta}_{++}$- and $\tilde{\theta}_{+-}$-directions are $2/(1 - (k_1/k_2))^{1/2}$ and $2/(1 + (k_1/k_2))^{1/2}$, respectively. Fig. 1 shows a two-soliton solution and Fig. 2 shows its source, $\tau_B$. The constant shift, $\alpha$, was assigned a non-traditional value ($\alpha = 0$) so as to show that the solution looks very much like a traditional solution, despite the different behavior of these two solutions in the limit $k_2 \to k_1$.

### 2.2.4 Limit of $k_2 \to k_1$

The structure of the traditional $\tau_T$ of Eq. (13) forces the solution to be reduced to a single-soliton in the limit. In the single-parameter family, constructed in terms $\tau_E$ of Eqs. (20), the limit depends on the value of $\alpha$. If $\alpha$ has the traditional singular wave number dependence of Eq. (17), then the limit of one soliton is attained. If $\alpha$ does not have this singular nature, the limit vanishes:

$$u(t,x)\Big|_{k_2 \to k_1} = \begin{cases} \dfrac{2k_1^2}{\cosh(k_1 x + 4k_1^3 t + \mu)^2} & \sinh\alpha = -\dfrac{2k_1 k_2}{(k_2 - k_1)(k_1 + k_2)} \\ 0 & \alpha = Const \end{cases} \tag{22}$$

## 2.3 Three-soliton solution

The traditional three-soliton $\tau$-function (wave numbers $k_3 > k_2 > k_1 > 0$) is:

$$\tau_T(t,x) = 1 + e^{2\theta_1} + e^{2\theta_2} + e^{2\theta_3} +$$
$$\left(\frac{k_2 - k_1}{k_1 + k_2}\right)^2 e^{2\theta_1} e^{2\theta_2} + \left(\frac{k_3 - k_1}{k_1 + k_3}\right)^2 e^{2\theta_1} e^{2\theta_3} + \left(\frac{k_3 - k_2}{k_2 + k_3}\right)^2 e^{2\theta_2} e^{2\theta_3} + \qquad (23)$$
$$\left(\frac{k_2 - k_1}{k_1 + k_2}\right)^2 \left(\frac{k_3 - k_1}{k_1 + k_3}\right)^2 \left(\frac{k_3 - k_2}{k_2 + k_3}\right)^2 e^{2\theta_1} e^{2\theta_2} e^{2\theta_3}$$

### 2.3.1 Three-parameter family of solutions

Following the two-soliton case, let us multiply Eq. (23) by

$$(k_1 + k_2)^2 (k_1 + k_3)^2 (k_2 + k_3)^2 e^{-\theta_1} e^{-\theta_2} e^{-\theta_3} .$$

Rearranging terms and eliminating an overall constant multiplicative factor, yields an equivalent $\tau$-function, which generates the same three-soliton solution as $\tau_T$ of Eq. Eq. (23):

$$\tau_E(t,x) = (k_2 - k_1)(k_3 - k_1)(k_3 - k_2)\cosh(\theta_{+++} + \alpha_{+++}) +$$
$$(k_2 - k_1)(k_1 + k_3)(k_2 + k_3)\cosh(\theta_{++-} + \alpha_{++-}) +$$
$$(k_1 + k_2)(k_3 - k_1)(k_2 + k_3)\cosh(\theta_{+-+} + \alpha_{+-+}) + \qquad (24)$$
$$(k_1 + k_2)(k_1 + k_3)(k_3 - k_2)\cosh(\theta_{+--} + \alpha_{+--})$$

In Eq. (24),

$$\theta_{\sigma_1\sigma_2\sigma_3} = \sum_{i=1}^{3} \sigma_i \theta_i \quad , \quad (\sigma_i = \pm 1) \quad , \qquad (25)$$

and $\theta_i$ are defined by Eq. (4)

The transformation from $\tau_T$ of Eq. (23) to $\tau_E$ dictates the values of the constant shifts, $\alpha_{\sigma_1\sigma_2\sigma_3}$, of Eq. (24) to be:

$$\sinh\alpha_{+++} = -2\frac{\left(k_1^2 k_2 + k_1 k_3^2 + k_2^2 k_3 + k_1 k_2 k_3\right)\left(k_1 k_2^2 + k_1^2 k_3 + k_2 k_3^2 + k_1 k_2 k_3\right)}{\left(k_2^2 - k_1^2\right)\left(k_3^2 - k_1^2\right)\left(k_3^2 - k_2^2\right)}$$
$$\sinh\alpha_{++-} = -\frac{2k_1 k_2}{\left(k_2^2 - k_1^2\right)} \quad , \quad \sinh\alpha_{+-+} = -\frac{2k_1 k_3}{\left(k_3^2 - k_1^2\right)} \quad , \quad \sinh\alpha_{+--} = -\frac{2k_2 k_3}{\left(k_3^2 - k_2^2\right)} \qquad (26)$$

However, using Eq. (24) in Eq. (2) yields a three-soliton solution of the KdV equation for *any* values of $\alpha_{\sigma_1\sigma_2\sigma_3}$, provided $\omega_i$ are given by Eq. (5), and $\alpha_{\sigma_1\sigma_2\sigma_3}$ obey the constraint:

$$\alpha_{+++} + \alpha_{+--} = \alpha_{++-} + \alpha_{+-+} . \tag{27}$$

(Clearly, the traditional $\alpha_{\sigma_1\sigma_2\sigma_3}$ of Eq. (26), obey Eq. (27).)

Thus, only three of the $\alpha_{\sigma_1\sigma_2\sigma_3}$ are linearly independent; the existence of a three-parameter family of three-soliton solutions has been established, of which the traditional solution is a member.

Finally, the wave-number dependent coefficients that multiply the hyperbolic cosines in Eq. (24) follow the general rule of Eq. (6). The product ($\sigma_i \sigma_j$) determines the sign within each coefficient.

**2.3.2 Solution parameters as soliton trajectory shifts**

The constraint of Eq. (27) is identical in shape to the constraint obeyed by the four $\theta_{\sigma_1\sigma_2\sigma_3}$'s because the latter are not linearly independent; they are constructed out of three independent $\theta_i$:

$$\theta_{\sigma_1\sigma_2\sigma_3} = \sum_{i=}^{3} \sigma_i \theta_i , \quad (\sigma_i = \pm 1) . \tag{28}$$

Eq. (28) leads to the constraint

$$\theta_{+++} + \theta_{+--} = \theta_{++-} + \theta_{+-+} . \tag{29}$$

This suggests a similar decomposition for $\alpha_{\sigma_1\sigma_2\sigma_3}$:

$$\alpha_{\sigma_1\sigma_2\sigma_3}(\vec{k}) = \sum_{i=1}^{3} \sigma_i \alpha_i(\vec{k}) . \tag{30}$$

Eq. (24) may be re-written in a form that exhibits the role of the shifts in soliton trajectories as solution parameters:

$$\begin{aligned}
\tau_E(t,x) = & (k_2 - k_1)(k_3 - k_1)(k_3 - k_2)\cosh\tilde{\theta}_{+++} + \\
& (k_2 - k_1)(k_1 + k_3)(k_2 + k_3)\cosh\tilde{\theta}_{++-} + \\
& (k_1 + k_2)(k_3 - k_1)(k_2 + k_3)\cosh\tilde{\theta}_{+-+} + \\
& (k_1 + k_2)(k_1 + k_3)(k_3 - k_2)\cosh\tilde{\theta}_{+--}
\end{aligned} \tag{31}$$

with

$$\tilde{\theta}_{\sigma_1\sigma_2\sigma_3\sigma_4} = \sum_{i=}^{4} \sigma_i \tilde{\theta}_i , \quad (\sigma_i = \pm 1) , \tag{32}$$

$$\tilde{\theta}_i = \theta_i + \varepsilon_i = k_i x + \omega_i t + \Delta_i(\vec{k}) , \quad (\Delta_i(\vec{k}) = \delta_i + \alpha_i(\vec{k})). \tag{33}$$

$\theta_i$ are defined in Eq. (4).

Thus, again, the parameters, on which the solution depends, have been formulated as shifts of soliton trajectories in the *x-t* plane. In the traditional case, in Eq. (31), $\alpha_i$ must contain wave number dependent contributions so that $\alpha_{\sigma_1 \sigma_2 \sigma_3}$ are given by Eq. (26).

### 2.3.3 Localized source
Following Eq. (7), define a new $\tau$-function based on Eq. (24):

$$\tau_B(t,x) = \frac{1}{\tau_E(t,x)} \ . \tag{34}$$

Eq. (8) generates the same three-parameter family of three-soliton solutions.

Unlike $\tau_T$ of Eq. (23) and $\tau_E$ of Eq. (31), $\tau_B$ is bounded in the *x-t* plane. Hence, under Eq. (8), the three-soliton solution is the image of a localized source. Figs. 3 and 4 show, respectively, a three-soliton solution and its source, $\tau_B$. Non-traditional values have been assigned to the shifts, $\alpha_{\sigma_1 \sigma_2 \sigma_3}$ ($\alpha_{\sigma_1 \sigma_2 \sigma_3} = 0$), so as to show that the solution looks very much like a traditional solution, despite the different behavior of these two solutions in the limit $k_2 \to k_1$.

### 2.3.2 Limit of $k_2 \to k_1$
When two wave numbers coincide, say, $k_2 \to k_1$, Eq. (23) becomes a traditional two-soliton $\tau$-function; the three-soliton solution is reduced to a two-soliton one. Using $\tau_E$ of Eq. (24) in the construction of the solution, the behavior of the latter in the limit depends on $\alpha_{\sigma_1 \sigma_2 \sigma_3}$. In the traditional case, the two-soliton limit is ensured owing to the singular nature of $\alpha_{\sigma_1 \sigma_2 \sigma_3}$ of Eq. (26). However, when $\alpha_{\sigma_1 \sigma_2 \sigma_3}$ do not obey Eq. (26), the limit is different. For example, if they are constants, the $k_2 \to k_1$ limit of the solution is a single-soliton solution. Examining Eq. (24) with $\alpha_{\sigma_1 \sigma_2 \sigma_3}$ independent of the wave numbers, one finds:

$$\tau_E(t,x)\big|_{k_2 \to k_1} = \left(4k_1\left(k_3^2 - k_1^2\right)\cosh\mu\right)\cosh(\theta_3 + v)$$

$$\mu = \delta_1 - \delta_2 + \frac{1}{2}(\alpha_{+-+} + \alpha_{+--}) \quad , \quad v = \delta_1 - \delta_2 + \frac{1}{2}(\alpha_{+-+} - \alpha_{+--})$$

(35)

Eq. (35) is a $\tau$-function that generates a single KdV-soliton solution, with wave number $k_3$.

### 2.4 Four-soliton solution

The traditional $\tau$-function for four-soliton solution (wave numbers $k_4 > k_3 > k_2 > k_1 > 0$) is

$$\tau_T(t,x) = 1 + \sum_{i=1}^{4} e^{2\theta_i} + \sum_{i=1}^{4}\sum_{j=i+1}^{4} \left(\frac{k_j - k_i}{k_i + k_j}\right)^2 e^{2\theta_i} e^{2\theta_j} +$$

$$\sum_{i=1}^{4}\sum_{j=i+1}^{4}\sum_{m=j+1}^{4} \left(\frac{k_j - k_i}{k_i + k_j}\right)^2 \left(\frac{k_m - k_i}{k_i + k_m}\right)^2 \left(\frac{k_m - k_j}{k_j + k_m}\right)^2 \left(\frac{k_j - k_i}{k_i + k_j}\right)^2 e^{2\theta_i} e^{2\theta_j} e^{2\theta_m} +$$

$$\left(\frac{k_2 - k_1}{k_1 + k_2}\right)^2 \left(\frac{k_3 - k_1}{k_1 + k_3}\right)^2 \left(\frac{k_4 - k_1}{k_1 + k_4}\right)^2 \left(\frac{k_3 - k_2}{k_2 + k_3}\right)^2 \left(\frac{k_4 - k_2}{k_4 + k_2}\right)^2 \left(\frac{k_4 - k_3}{k_3 + k_4}\right)^2 e^{2\theta_1} e^{2\theta_2} e^{2\theta_3} e^{2\theta_4}$$

(36)

#### 2.4.1 Four-parameter family of solutions

Following the procedure delineated in Appendix I, one obtains an equivalent $\tau$-function:

$$\begin{aligned}
\tau_E(t,x) = &\,(k_2 - k_1)(k_3 - k_1)(k_4 - k_1)(k_3 - k_2)(k_4 - k_2)(k_4 - k_3)\cosh(\theta_{++++} + \alpha_{++++}) + \\
&\,(k_2 - k_1)(k_3 - k_1)(k_4 + k_1)(k_3 - k_2)(k_4 + k_2)(k_4 + k_3)\cosh(\theta_{+++-} + \alpha_{+++-}) + \\
&\,(k_2 - k_1)(k_3 + k_1)(k_4 - k_1)(k_3 + k_2)(k_4 - k_2)(k_4 + k_3)\cosh(\theta_{++-+} + \alpha_{++-+}) + \\
&\,(k_2 + k_1)(k_3 - k_1)(k_4 - k_1)(k_3 + k_2)(k_4 + k_2)(k_4 - k_3)\cosh(\theta_{+-++} + \alpha_{+-++}) + \\
&\,(k_2 - k_1)(k_3 + k_1)(k_4 + k_1)(k_3 + k_2)(k_4 + k_2)(k_4 - k_3)\cosh(\theta_{++--} + \alpha_{++--}) + \\
&\,(k_2 + k_1)(k_3 - k_1)(k_4 + k_1)(k_3 + k_2)(k_4 - k_2)(k_4 + k_3)\cosh(\theta_{+-+-} + \alpha_{+-+-}) + \\
&\,(k_2 + k_1)(k_3 + k_1)(k_4 - k_1)(k_3 - k_2)(k_4 + k_2)(k_4 + k_3)\cosh(\theta_{+--+} + \alpha_{+--+}) + \\
&\,(k_2 + k_1)(k_3 + k_1)(k_4 + k_1)(k_3 - k_2)(k_4 - k_2)(k_4 - k_3)\cosh(\theta_{+---} + \alpha_{+---})
\end{aligned}$$

(37)

$$\theta_{\vec{\sigma}} = \sum_{i=1}^{4} \sigma_i \theta_i$$

$$\left(\vec{\sigma} = \{\sigma_1, \sigma_2, \sigma_3, \sigma_4\} \quad , \quad \sigma_1 = +1, \sigma_{i>1} = \pm 1\right)$$

(38)

$\theta_i$ are defined in Eq. (4) and $\alpha_{\vec{\sigma}}$. Following the procedure delineated in Appendix I, the structure of $\tau_T$ of Eq. (36) dictates the expressions for $\alpha_{\vec{\sigma}}$ in the traditional case. However, using Eq. (37)

in Eq. (2) yields a four-soliton solution of the KdV equation for *any* values of $\alpha_{\vec{\sigma}}$, provided $\omega_i$ are given by Eq. (5), and $\alpha_{\vec{\sigma}}$ obey the constraints:

$$\begin{aligned}
\alpha_{++--} &= \alpha_{+++-} - \alpha_{+-++} + \alpha_{+--+} \quad, \quad \alpha_{+--+} = \alpha_{++-+} - \alpha_{+++-} + \alpha_{+-+-} \\
\alpha_{+---} &= \alpha_{++--} + \alpha_{+-+-} - \alpha_{+++-} \quad, \quad \alpha_{+-+-} = \alpha_{+--+} - \alpha_{++--} + 2\alpha_{+++-} - \alpha_{++++}
\end{aligned} \quad (39)$$

Thus, the existence of a four-parameter family of four-soliton solutions has been established, of which the traditional solution is just one member.

### 2.4.2 Solution parameters as soliton trajectory shifts

As in the case of the three-soliton solution, the constraints of Eq. (39) are identical in shape to four constraints obeyed by $\theta_{\vec{\sigma}}$. The latter are a trivial expression of the fact that the eight $\theta_{\vec{\sigma}}$ are linear combinations of only four independent $\theta_i$ (see Eq. (6)). This allows, again, for the construction of the eight linearly dependent shifts in terms of four independent shifts:

$$\alpha_{\vec{\sigma}}(\vec{k}) = \sum_{i=1}^{4} \sigma_i \alpha_i(\vec{k}) \quad , \quad (\vec{k} = \{k_1, k_2, k_3, k_4\}) \quad (40)$$

and for re-writing of Eq. (37) in the form of Eq. (6). Again, the parameters, on which the four-soliton solution depends, have been formulated as shifts of soliton trajectories in the *x-t* plane.

### 2.4.3 Limit of coinciding wave numbers

The structure of $\tau_T$ of Eq. (36) ensures that, when $k_2 \to k_1$, the solution is reduced to a three-soliton solution (wave numbers $k_1$, $k_3$ and $k_4$). If one next considers the limit of $k_4 \to k_3$, then the three-soliton solution is reduced to a two-soliton solution (wave numbers $k_1$ and $k_3$). If the solution is constructed from $\tau_E$ of Eq. (37), the result in the limit depends on $\alpha_{\vec{\sigma}}$. If the latter assume the singular wave-number dependent values dictated by the structure of $\tau_T$, then the traditional limit is reached. However, if $\alpha_{\vec{\sigma}}$ do not assume the traditional values, the limit may be different. For example, if all $\alpha_{\vec{\sigma}}$ are constants, the limit, $k_2 \to k_1$, of the four-soliton solution is a two-soliton solu-tion (wave numbers $k_3$ and $k_4$). Imposing, in addition, $k_4 \to k_3$, this two-soliton solution is reduced to zero.

### 2.4.3 Localized source
Finally, use Eq. (37) in the definition, Eq. (34), of $\tau_B$. Under Eq. (8), the four-soliton solution is the image of a source that is localized in the *x-t* plane.

## 3. The Kadomtsev-Petviashvili II equation
The line-soliton solutions of the Kadomtsev-Petviashvili II (KP II) equation,

$$\frac{\partial}{\partial x}\left(-4\frac{\partial u}{\partial t}+\frac{\partial^3 u}{\partial x^3}+6u\frac{\partial u}{\partial x}\right)+3\frac{\partial^2 u}{\partial y^2}=0 ,  \tag{41}$$

are constructed as follows [17, 18]:

$$u(t,x,y)=2\partial_x^2 \log\{\tau(t,x,y)\} .  \tag{42}$$

The traditional $\tau$-function is given by

$$\tau_T(t,x,y)=\begin{cases} \sum_{i=1}^{M}\xi_M(i)e^{\theta_i(t,x,y)} & N=1 \\ \sum_{i=1}^{M}\xi_M(i)e^{\sum_{j=1,j\neq i}^{M}\theta_j(t,x,y)} & N=M-1 \\ \sum_{1\leq i_1<...<i_N\leq M}\xi_M(i_1,...,i_N)\left(\prod_{1\leq j<l\leq N}(k_{i_l}-k_{i_j})\right)e^{\sum_{j=1}^{N}\theta_{i_j}(t,x,y)} & 2\leq N\leq M-2 \end{cases}, \tag{43}$$

$$k_1<k_2<...<k_M , \tag{44}$$

$$\theta_i(t,x,y)=-k_i x + k_i^2 y - k_i^3 t . \tag{45}$$

In Eqs. (43) and (44), $M$ is the size of a set of wave numbers, $\{k_1,...,k_M\}$. The sum goes over all $\binom{M}{N}$ subsets of $N$ wave numbers.

To exclude singular solutions of Eq. (41), one requires

$$\xi_M(i_1,...,i_N)\geq 0 . \tag{46}$$

Apart from positivity, the coefficients, $\xi_M(i)$, with $N = 1$ and $N = M-1$, may assume arbitrary values. For $2 \leq N \leq M-2$, $\xi_M(i_1,....,i_N)$ are constrained by the Plücker relations (see, E.g. [19]). For example, for $(M,N) = (4,2)$ one finds a single Plücker relation:

$$\xi_4(1,2)\xi_4(3,4) - \xi_4(1,3)\xi_4(2,4) + \xi_4(1,4)\xi_4(2,3) = 0 \ . \tag{47}$$

### 3.1 Generating a bounded $\tau_B$-function

To generate a bounded $\tau$-function that is localized in space through the recipe of

$$\tau_B(t,x) = \frac{1}{\tau_E(t,x,y)} \ , \tag{48}$$

one needs to ensure that $\tau_E(t,x,y)$ does not vanish asymptotically in some domain in the (1+2)-dimensional space. To avoid this, let us replace $\tau_T$ of Eq. (43) by an equivalent $\tau$-function:

$$\tau_E(t,x,y) = e^{-\mu \sum_{i=1}^{M} \theta_i(t,x,y)} \tau_T(t,x,y) \ , \tag{49}$$

and use Eq. (49) in Eq. (42). The multiplicative factor does not change the soliton solution.

Now, using Eq. (43), write Eq. (49) as a sum of exponentials. The generic form of the exponential terms in the result is:

$$e^{\left(-\mu \sum_{i=1}^{M} \theta_i(t,x,y) + \sum_{j=1}^{N} \theta_{i_j}(t,x,y)\right)} \ . \tag{50}$$

To ensure that $\tau_B$ of Eq. (48) is localized in space, one must ensure that that not all such exponential terms vanish simultaneously in some domain in the (1+2)-dimensional space. This requires that the exponents in the exponential terms of the type of Eq. (50) do not all become indefinitely large and negative simultaneously in some domain; some exponents must become large and positive. This can be achieved by requiring that the sum of the exponents in all the terms in $\tau_E$ of Eq. (49) vanishes. In that sum, each $\theta_i(t,x,y)$ is multiplied by

$$(1-\mu)\binom{M-1}{N-1} - \mu\left(\binom{M}{N} - \binom{M-1}{N-1}\right). \tag{51}$$

The first term in Eq. (51) counts the number of times each $\theta_i(t,x,y)$ appears with a positive sign, and the second term counts the number of times it appear with a negative sign. As all the $\theta_i(t,x,y)$ are independent, the vanishing of the sum requires that the coefficient of each $\theta_i(t,x,y)$ vanish:

$$\mu = \binom{M-1}{N-1} \bigg/ \binom{M}{N}. \tag{52}$$

With this choice, $\tau_B$ is localized in the (1+2)-dimensional space, a hump in the $x$-$y$ plane at all times, which, through Eq. (8), serves as a localized source for a solution of Eq. (41).

**3.2 ($M \geq 2$, $N = 1$)-solutions**

For such solutions, Eq. (52) yields $\mu = (1/M)$. By Eq. (43), these solutions have no wave-number dependent coefficients. Hence, they cannot be extended to multi-parameter families of solutions. However, making the solutions images of localized sources is possible. Here are some examples.

**3.2.1 Two wave numbers: Single-soliton solution**

The single-soliton solution is constructed from

$$\tau_S(t,x,y) = \xi_1 \exp(\theta_1(t,x,y)) + \xi_2 \exp(\theta_2(t,x,y)) . \tag{53}$$

With $M = 2$, $N = 1$, Eq. (52) requires $\mu = (1/2)$, yielding (eliminating a constant multiplicative factor that does not affect the soliton solution):

$$\tau_B(t,x,y) = \frac{1}{\cosh\left(\frac{1}{2}(\theta_1(t,x,y) - \theta_2(t,x,y)) + \delta\right)} , \quad \left(\delta = \operatorname{arctanh}\left(\frac{\xi_1 - \xi_2}{\xi_1 + \xi_2}\right)\right), \tag{54}$$

which, through

$$u(t,x,y) = -2\partial_x^2 \log \tau_B . \tag{55}$$

generates the same single-soliton solution. Thus, as in the case of the KdV equation, the single-soliton solution is the image of the single soliton given by Eq. (54).

**3.2.2 Three wave numbers: Three-soliton solution ($M = 3$, $N = 1$)**

The traditional $\tau$-function for the three-soliton solution (Y-shaped solution) is:

$$\tau_T(t,x,y) = \xi_1 \exp(\theta_1(t,x,y)) + \xi_2 \exp(\theta_2(t,x,y)) + \xi_3 \exp(\theta_3(t,x,y)) . \tag{56}$$

This solution propagates rigidly in the x-y plane with a velocity given by [20]

$$\begin{aligned} v_x &= k_1 k_2 + k_1 k_3 + k_2 k_3 \\ v_y &= k_1 + k_2 + k_3 \end{aligned}. \qquad (57)$$

Eq. (52) requires $\mu = (1/3)$. The resulting equivalent $\tau$-function is:

$$\begin{aligned} \tau_E(t,x,y) &= \xi_1 \exp\left(\frac{1}{3}(2\theta_1(t,x,y) - \theta_2(t,x,y) - \theta_3(t,x,y))\right) + \\ &\quad \xi_2 \exp\left(\frac{1}{3}(2\theta_2(t,x,y) - \theta_1(t,x,y) - \theta_3(t,x,y))\right) + \quad , \qquad (58) \\ &\quad \xi_3 \exp\left(\left(\frac{1}{3}(2\theta_3(t,x,y) - \theta_1(t,x,y) - \theta_2(t,x,y))\right)\right) \end{aligned}$$

$\tau_E$ of Eq. (58) generates the same three-soliton solution as $\tau_T$ of Eq. (56). Let us now use it in the definition of $\tau_B$ in Eq. (48). $\tau_B$ describes a hump that is localized in the x-y plane at any time. The position of its maximum is located at the point in the plane, for which

$$\frac{1}{3}\log \xi_1 + \theta_1(t,x,y) = \frac{1}{3}\log \xi_2 + \theta_2(t,x,y) = \frac{1}{3}\log \xi_3 + \theta_3(t,x,y) \quad . \qquad (59)$$

Eq. (59) yields the coordinates x and y of the point of maximum as functions of t. The velocity of propagation of the source is computed to be the velocity of the solution, given in Eq. (57). Figs. 5 and 6 show, respectively, a three-soliton solution and its localized source.

### 3.2.3 Four wave numbers: (4,1) Four-soliton solution
The traditional $\tau$-function for the four-soliton solution with ($M=4$, $N=1$) is:

$$\tau_T(t,x,y) = \xi_1 \exp(\theta_1(t,x,y)) + \xi_2 \exp(\theta_2(t,x,y)) + \xi_3 \exp(\theta_3(t,x,y)) + \xi_4 \exp(\theta_4(t,x,y)) \quad . \qquad (60)$$

To obtain a localized $\tau_B$, Eq. (52) requires $\mu = (1/4)$. Figs. 7 and 8 show, respectively, a four-soliton solution and its localized source, $\tau_B$.

### 3.3 Four wave numbers: (4,2) Four-soliton solution
In (M,N) solutions with $N > 1$, the numerical coefficients depend on the wave numbers (see Eq. (43), allowing for the extension of the multi-soliton solutions to a multi-parameter family of solu-tions, the traditional solutions being just a subset of this family. An elegant algorithm for the procedure has been found only in the case of solutions with ($M = 2k$, $N = k > 1$). The case of the (4,2) solution is discussed as an example. The traditional $\tau$-function is given by

$$\tau_T = \xi_4(1,2)(k_2-k_1)e^{\theta_1+\theta_2} + \xi_4(1,3)(k_3-k_1)e^{\theta_1+\theta_3} + \xi_4(1,4)(k_4-k_1)e^{\theta_1+\theta_4} \\ \xi_4(2,3)(k_3-k_2)e^{\theta_2+\theta_3} + \xi_4(2,4)(k_4-k_2)e^{\theta_2+\theta_4} + \xi_4(3,4)(k_4-k_3)e^{\theta_3+\theta_4}. \tag{61}$$

In Eq. (61), $(t,x,y)$, have been omitted from $\theta_i$ for the sake of brevity. The $\xi$'s obey Eq. (47).

### 3.3.1 Three-parameter family of solutions
Eq. (52) requires $\mu = (1/2)$. Regrouping terms, the resulting equivalent $\tau$-function is rewritten as:

$$\tau_E(t,x,y) = \sqrt{(k_2-k_1)(k_4-k_3)\xi_4(1,2)\xi_4(3,4)}\cosh\left\{\frac{1}{2}\theta_{++--} + \alpha_{++--}\right\} + \\ \sqrt{(k_3-k_1)(k_4-k_2)\xi_4(1,3)\xi_4(2,4)}\cosh\left\{\frac{1}{2}\theta_{+-+-} + \alpha_{+-+-}\right\} + \tag{62} \\ \sqrt{(k_3-k_2)(k_4-k_1)\xi_4(1,4)\xi_4(2,3)}\cosh\left\{\frac{1}{2}\theta_{+--+} + \alpha_{+--+}\right\}$$

$$\theta_{\sigma_1\sigma_2\sigma_3\sigma_4} = \sigma_1\theta_1 + \sigma_2\theta_2 + \sigma_3\theta_3 + \sigma_4\theta_4, \tag{63}$$

$$\alpha_{++--} = \log\left\{\sqrt{\frac{(k_2-k_1)\xi_4(1,2)}{(k_4-k_3)\xi_4(3,4)}}\right\}, \quad \alpha_{+-+-} = \log\left\{\sqrt{\frac{(k_3-k_1)\xi_4(1,3)}{(k_4-k_2)\xi_4(2,4)}}\right\} \\ \alpha_{+--+} = \log\left\{\sqrt{\frac{(k_4-k_1)\xi_4(1,4)}{(k_3-k_2)\xi_4(2,3)}}\right\} \tag{64}$$

(The notation is as in the case of the KdV equation. A constant multiplicative factor has been removed from $\tau_E$.)

With $\alpha_{\vec{\sigma}}$ of Eq. (64), $\tau_E$-of Eq. (62) generates the same (4,2) solution as $\tau_T$ of Eq. (61). However, substituting $\tau_E$ in Eq. (42), one finds that it generates a (4,2) solution of the KP II equation for *any* values of $\alpha_{\vec{\sigma}}$. Thus, the existence of a three-parameter family of (4,2) solutions has been established, the traditional solution being just one member of it.

### 3.3.2 Limit of $k_2 \to k_1$

Consider now the limit when two wave numbers coincide, say, $k_2 \to k_1$. From $\tau_T$ of Eq. (61) one deduces that the (4,2) solution is reduced to a (3,2) solution, which is a three-soliton solution (Y-shaped) with wave numbers $k_1$, $k_3$ and $k_4$. Constructing the solution through Eq. (62), this limit is reached if $\alpha_{\vec{\sigma}}$ are assigned the required traditional singular expressions of Eq. (64)). If they have other values, then the $k_2 \to k_1$ limit may be different. For constant $\alpha_{\vec{\sigma}}$, $\tau_E$ of Eq. (62) tends to:

$$\tau_E(t,x,y)\Big|_{k_2 \to k_1} = \sqrt{(k_3 - k_1)(k_4 - k_1)} \times$$
$$\left( \begin{array}{l} \sqrt{\xi_4(1,3)\xi_4(2,4)} \cosh\left\{ \frac{1}{2}\left( -(k_3 - k_4)x + (k_3^2 - k_4^2)y - (k_3^3 - k_4^3)t \right) + \alpha_{+-+-} \right\} + \\ \sqrt{\xi_4(1,4)\xi_4(2,3)} \cosh\left\{ \frac{1}{2}\left( -(k_3 - k_4)x + (k_3^2 - k_4^2)y - (k_3^3 - k_4^3)t \right) - \alpha_{+--+} \right\} \end{array} \right), \quad (65)$$

which generates a single-soliton solution with wave numbers $k_3$ and $k_4$!

### 3.3.3 Localized source

Clearly, using Eq. (62) in the definition of $\tau_B$ by Eq. (48), Eq. (55) generates the same family of (4,2) solutions. However, unlike $\tau_T$ and $\tau_E$, $\tau_B$ is bounded in the whole (1+2)-dimensional space, and generates a source that is localized in the *x-y* plane. The (4,2) solution is the image of this localized under Eq. (55). Figs. 9 and 10 present a (4,2) solution and its source, $\tau_B$, respectively. The $\alpha_{\vec{\sigma}}$ have been assigned non-traditional values (all = 0) to show that despite the difference in properties of the traditional and the new solutions they look very much alike.

## 4. The modified KdV equation

The soliton solutions of the modified KdV (mKdV) equation,

$$u_t = 6u^2 u_x + u_{xxx}, \quad (66)$$

are constructed through a transformation of a different structure [15]:

$$u(t,x) = 2\partial_x \arctan(\tau(t,x)). \quad (67)$$

Owing to the fact that the connection between the solution and the $\tau$-function is not through a logarithmic transformation, a simple procedure of the type described in Sections 2 and 3 has not been found. However, the Miura transformation connecting the solutions of the KdV and mKdV equations [21] ensures that the extension of the traditional mKdV-soliton solutions to a wider family of solutions, which depend on free shifts as parameters, is possible here as well. Rather than embarking upon a full analysis, let us present here the case of the two-soliton solution. For the latter, the traditional $\tau$-function is given by:

$$\tau_T(t,x) = \frac{e^{\theta_1} + e^{\theta_2}}{1 - \left(\frac{k_2 - k_1}{k_1 + k_2}\right)^2 e^{\theta_1} e^{\theta_2}} \qquad (68)$$

$$\left(\theta_i = k_i x + k_i^3 t + \delta_i\right)$$

$\tau_T$ of Eq. (68) is unbounded in the vicinity of a line in the $x$-$t$ plane. This singular behavior is of no concern, as it is remedied by the transformation in Eq. (67).

To expose the existence a single-parameter family of two-soliton solutions, multiply the top and the bottom of Eq. (68) by

$$\left(k_1 + k_2\right)^2 e^{-\frac{1}{2}(\theta_1 + \theta_2)}. \qquad (69)$$

The result leads to the following equivalent $\tau$-function:

$$\tau_E(t,x) = -\frac{k_1 + k_2}{k_2 - k_1} \frac{\cosh\left(\frac{1}{2}(\theta_1 - \theta_2)\right)}{\sinh\left(\frac{1}{2}(\theta_1 + \theta_2) + \alpha\right)} . \qquad (70)$$

The structure of $\tau_T$ of Eq. (68) dictates the wave-number dependence of $\alpha$ to be:

$$\sinh \alpha = -\frac{2 k_1 k_2}{k_2^2 - k_1^2} . \qquad (71)$$

Substituting Eq. (70) in Eq. (67), one finds that $u(t,x)$ is a two-soliton solution of Eq. (65), indeed, for *any* value of $\alpha$. Thus, the existence of a single-parameter family of two-soliton solutions of the mKdV equation has been established; the traditional solution is one member in this family.

As evident from Eq. (68), in the traditional case, the two soliton solution is reduced to a single-soliton solution when $k_2 \rightarrow k_1$. Using Eq. (70), this is a consequence of the singular nature of the traditional value of $\alpha$, given by Eq. (71). However, if $\alpha$ does not have the singular structure of Eq. (71), the limit is different. The leading singular term in Eq. (70) is then:

$$\tau(t,x)\Big|_{k_2 \rightarrow k_1} \sim -\frac{2k_1}{k_2 - k_1} \frac{\cosh\left(\frac{1}{2}(\delta_1 - \delta_2)\right)}{\sinh\left(k_1 x + k_1^3 t + \frac{\delta_1 + \delta_2}{2} + \alpha\right)} \xrightarrow[k_2 \rightarrow k_1]{} \pm \infty \ , \tag{72}$$

where the final sign depends on which side of the line

$$k_1 x + k_1^3 t + \frac{\delta_1 + \delta_2}{2} + \alpha = 0 \tag{73}$$

one is. Hence, in the limit, the ArcTanh in Eq. (67) jumps between $-\pi/2$ and $+\pi/2$. Consequently, the limit of the two-soliton solution is a zero-width single soliton:

$$u(t,x) \xrightarrow[k_2 \rightarrow k_1]{} 2\pi \delta\left(k_1 x + k_1^3 t + \frac{\delta_1 + \delta_2}{2} + \alpha\right) . \tag{74}$$

## 5. Concluding comments

In this paper it has been shown that known evolution equations have parametric families of multi-soliton solutions that are far wider than the solutions constructed in the traditional Inverse-Scattering/Hirota approach. The traditional solutions are just a subset within this family. While the solutions in the traditional set and the ones outside the set look very much alike, their characteristics may be quite different. Here, it has been shown that when two wave numbers coincide in a multi-soliton solution, the limits for solutions within the traditional subset and outside this subset are markedly different. In the case of the KdV equation, a traditional $N$-soliton solution degener-ates into a traditional ($N - 1$)-soliton solution, whereas outside the traditional subset, it may degenerate into an ($N - 2$)-soliton solution. In the case of the KP II equation, the (4,2) solution has been discussed. In the traditional case, it reduces into a three—soliton solution, whereas outside the traditional subset, it is reduced to a single-solution. In the case of the mKdV equation, the traditional two-soliton solution is reduced to a single-soliton solution, whereas outside the traditional subset, it may degenerate into a $\delta$-function.

Clearly, the analysis presented here can be applied to other KdV-like equations, such as the bi-directional KdV equation [22] and the Sawada-Kotera [23] equation. This, obviously opens the door to a far richer spectrum of soliton-solutions in the cases of other evolution equations.

**Appendix I. Construction of equivalent $\tau$-function for $N$-soliton solution of KdV equation**

One first multiplies the traditional $\tau$-function,

$$\tau_T(t,x) = 1 + \sum_{n=1}^{N} \sum_{1 \leq i_1 < i_2 < \cdots < i_n \leq N} \left( \prod_{j=1}^{n} \prod_{l=j+1}^{n} \left( \frac{k_{i_l} - k_{i_j}}{k_{i_l} + k_{i_j}} \right)^2 \right) e^{2\sum_{m=1}^{n} \theta_{i_m}}, \quad (I.1)$$

by

$$\left( \prod_{j=1}^{N} \prod_{l=j+1}^{N} (k_j + k_l)^2 \right) \prod_{i=1}^{N} e^{-\theta_i} \quad (I.2)$$

The $2^N$ exponential terms in the result,

$$\left( \prod_{j=1}^{N} \prod_{l=j+1}^{N} (k_j + k_l)^2 \right) \prod_{i=1}^{N} e^{-\theta_i} \left( 1 + \sum_{n=1}^{N} \sum_{1 \leq i_1 < i_2 < \cdots < i_n \leq N} \left( \prod_{j=1}^{n} \prod_{l=j+1}^{n} \left( \frac{k_{i_l} - k_{i_j}}{k_{i_l} + k_{i_j}} \right)^2 \right) e^{2\sum_{m=1}^{n} \theta_{i_m}} \right), \quad (I.3)$$

are split into $2^{N-1}$ pairs of terms. The simplest pair is the one, in which all $\theta_i$ have the same signs. It is obtained from the sum of the following two terms in $\tau_T$:

$$1 \quad , \quad \left( \prod_{j=1}^{N} \prod_{l=j+1}^{N} \left( \frac{k_{i_l} - k_{i_j}}{k_{i_l} + k_{i_j}} \right)^2 \right) e^{2\sum_{m=1}^{N} \theta_{i_m}}. \quad (I.4)$$

When these are multiplied by the factor of Eq. (I.2), the sum of the two terms becomes

$$\left(\prod_{j=1}^{N}\prod_{l=j+1}^{N}(k_j+k_l)^2\right)e^{-\sum_{i=1}^{N}\theta_i} + \left(\prod_{j=1}^{N}\prod_{l=j+1}^{N}(k_j-k_l)^2\right)e^{\sum_{i=1}^{N}\theta_i} =$$

$$\left\{\prod_{j=1}^{N}\prod_{l=j+1}^{N}(k_j+k_l)^2 + \prod_{j=1}^{N}\prod_{l=j+1}^{N}(k_j-k_l)^2\right\}\cos\theta_{\underbrace{++\cdots+}_{N\text{ times}}} + \quad . \qquad (I.5)$$

$$\left\{\prod_{j=1}^{N}\prod_{l=j+1}^{N}(k_j-k_l)^2 - \prod_{j=1}^{N}\prod_{l=j+1}^{N}(k_j+k_l)^2\right\}\sin\theta_{\underbrace{++\cdots+}_{N\text{ times}}}$$

This term can be re-written as:

$$\left(2\prod_{j=1}^{N}\prod_{l=j+1}^{N}(k_j+k_l)\right)\left(\prod_{j=1}^{N}\prod_{l=j+1}^{N}(k_j-k_l)\right)\cos\left(\theta_{\underbrace{++\cdots+}_{N\text{ times}}}+\alpha_{\underbrace{++\cdots+}_{N\text{ times}}}\right), \qquad (I.6)$$

where

$$\sinh\alpha_{\underbrace{++\cdots+}_{N\text{ times}}} = \frac{\prod_{j=1}^{N}\prod_{l=j+1}^{N}(k_j-k_l)^2 - \prod_{j=1}^{N}\prod_{l=j+1}^{N}(k_j+k_l)^2}{2\prod_{j=1}^{N}\prod_{l=j+1}^{N}(k_j+k_l)\prod_{j=1}^{N}\prod_{l=j+1}^{N}(k_j-k_l)} . \qquad (I.7)$$

The pair of the next level of complication, is that, in which one of the $\theta_i$ has a negative sign. Take as an example the case that this is $\theta_N$. It is obtained from the sum of the following two terms in $\tau_T$:

$$e^{2\theta_N} \quad , \quad \left(\prod_{j=1}^{N-1}\prod_{l=j+1}^{N-1}\left(\frac{k_{i_l}-k_{i_j}}{k_{i_l}+k_{i_j}}\right)^2\right)e^{2\sum_{m=1}^{N-1}\theta_{i_m}} . \qquad (I.8)$$

When these are multiplied by the factor of Eq. (I.2), the sum of the two terms becomes

$$\left(\prod_{j=1}^{N}\prod_{l=j+1}^{N}(k_j+k_l)^2\right)e^{-\sum_{i=1}^{N-1}\theta_i+\theta_N} + \prod_{i=1}^{N-1}(k_N+k_l)^2\left(\prod_{j=1}^{N-1}\prod_{l=j+1}^{N-1}(k_j-k_l)^2\right)e^{\sum_{i=1}^{N-1}\theta_i-\theta_N} =$$

$$\left\{\prod_{j=1}^{N}\prod_{l=j+1}^{N}(k_j+k_l)^2 + \prod_{i=1}^{N-1}(k_N+k_l)^2\left(\prod_{j=1}^{N-1}\prod_{l=j+1}^{N-1}(k_j-k_l)^2\right)\right\}\cos\theta_{\underbrace{++\cdots+-}_{N-1\text{ times}}} + \quad . \qquad (I.9)$$

$$\left\{\prod_{i=1}^{N-1}(k_N+k_l)^2\left(\prod_{j=1}^{N-1}\prod_{l=j+1}^{N-1}(k_j-k_l)^2\right) - \prod_{j=1}^{N}\prod_{l=j+1}^{N}(k_j-k_l)^2\right\}\sin\theta_{\underbrace{++\cdots+-}_{N-1\text{ times}}}$$

This term can be re-written as

$$\left(2\prod_{j=1}^{N}\prod_{l=j+1}^{N}(k_j+k_l)\right)\prod_{i=1}^{N-1}(k_N+k_l)\left(\prod_{j=1}^{N-1}\prod_{l=j+1}^{N-1}(k_j-k_l)\right)\cos\left(\theta_{\underbrace{++\cdots+-}_{N-1\text{ times}}}+\alpha_{\underbrace{++\cdots+-}_{N-1\text{ times}}}\right), \qquad (I.10)$$

where

$$\sinh\alpha_{\underbrace{++\cdots+}_{N-1\text{ times}}--} = \frac{\prod_{i=1}^{N-1}(k_N+k_i)^2\left(\prod_{j=1}^{N-1}\prod_{l=j+1}^{N-1}(k_j-k_l)^2\right) - \prod_{j=1}^{N}\prod_{l=j+1}^{N}(k_j+k_l)^2}{\left(2\prod_{j=1}^{N}\prod_{l=j+1}^{N}(k_j+k_l)\right)\left(\prod_{j=1}^{N}\prod_{l=j+1}^{N}(k_j-k_l)\right)} \quad . \tag{I.11}$$

The next level of complication is in pairs, in which two of the $\theta_i$ has a negative sign. Take as an example the case that these are $\theta_N$ and $\theta_{N-1}$. It is obtained from the sum of the following two terms in $\tau_T$:

$$e^{2(\theta_{N-1}+\theta_N)} \quad , \quad \left(\prod_{j=1}^{N-2}\prod_{l=j+1}^{N-2}\left(\frac{k_{i_l}-k_{i_j}}{k_{i_l}+k_{i_j}}\right)^2\right)e^{2\sum_{m=1}^{N-2}\theta_{i_m}} \quad . \tag{I.12}$$

When these are multiplied by the factor of Eq. (I.2), their sum becomes

$$\left(\prod_{j=1}^{N}\prod_{l=j+1}^{N}(k_j+k_l)^2\right)e^{-\sum_{i=1}^{N-2}\theta_i+\theta_{N-1}+\theta_N} + $$

$$\prod_{i=1}^{N-1}(k_N+k_i)^2\prod_{i=1}^{N-2}(k_{N-1}+k_i)^2\left(\prod_{j=1}^{N-2}\prod_{l=j+1}^{N-2}(k_j-k_l)^2\right)e^{\sum_{i=1}^{N-2}\theta_i-\theta_{N-1}-\theta_N} = $$

$$\left\{\prod_{j=1}^{N}\prod_{l=j+1}^{N}(k_j+k_l)^2+\prod_{i=1}^{N-1}(k_N+k_i)^2\prod_{i=1}^{N-2}(k_{N-1}+k_i)^2\left(\prod_{j=1}^{N-2}\prod_{l=j+1}^{N-2}(k_j-k_l)^2\right)\right\}\cos\theta_{\underbrace{++\cdots+}_{N-1\text{ times}}-} + $$

$$\left\{\prod_{i=1}^{N-1}(k_N+k_i)^2\prod_{i=1}^{N-2}(k_{N-1}+k_i)^2\left(\prod_{j=1}^{N-2}\prod_{l=j+1}^{N-2}(k_j-k_l)^2\right) - \prod_{j=1}^{N}\prod_{l=j+1}^{N}(k_j-k_l)^2\right\}\sin\theta_{\underbrace{++\cdots+}_{N-1\text{ times}}-}$$

. (I.13)

This term can be re-written as

$$\left(2\prod_{j=1}^{N}\prod_{l=j+1}^{N}(k_j+k_l)\right)\left(\prod_{i=1}^{N-1}(k_N+k_i)\prod_{i=1}^{N-2}(k_{N-1}+k_i)\left(\prod_{j=1}^{N-2}\prod_{l=j+1}^{N-2}(k_j-k_l)\right)\right)\times$$

$$\cos\left(\theta_{\underbrace{++\cdots+}_{N-2\text{ times}}--}+\alpha_{\underbrace{++\cdots+}_{N-2\text{ times}}--}\right) \quad , \tag{I.14}$$

where

$$\sinh\alpha_{\underbrace{++\cdots+}_{N-2\text{ times}}--} = \frac{\prod_{i=1}^{N-1}(k_N+k_i)^2\prod_{i=1}^{N-2}(k_{N-1}+k_i)^2\left(\prod_{j=1}^{N-2}\prod_{l=j+1}^{N-2}(k_j-k_l)^2\right) - \prod_{j=1}^{N}\prod_{l=j+1}^{N}(k_j+k_l)^2}{\left(2\prod_{j=1}^{N}\prod_{l=j+1}^{N}(k_j+k_l)\right)\left(\prod_{j=1}^{N}\prod_{l=j+1}^{N}(k_j-k_l)\right)} \quad . \tag{I.15}$$

The construction of the remaining part of $\tau_E$ follows similar steps. All other pairs of exponential terms in $\tau_T$ are treated in a similar manner.

Note that, in the three examples discussed above, the final form of each contribution is proportional to a constant multiplicative factor:

$$\left(2\prod_{j=1}^{N}\prod_{l=j+1}^{N}(k_j + k_l)\right) . \tag{I.16}$$

This factor emerges in the final form of the contributions of all $2^{N-1}$ pairs of terms. Hence, the final result is proportional to that factor. As such a factor in a $\tau$-function does not affect $u(t,x)$ given by Eq. (2), it is omitted in the final definition of the equivalent $\tau$-function, $\tau_E$. As a result, the contributions to $\tau_E$ of the three terms discussed in detail above become:

Eq. (I.6): $$\left(\prod_{j=1}^{N}\prod_{l=j+1}^{N}(k_j - k_l)\right)\cos\left(\theta_{\underbrace{++\cdots+}_{N \text{ times}}} + \alpha_{\underbrace{++\cdots+}_{N \text{ times}}}\right), \tag{I.17}$$

Eq. (I.10): $$\prod_{i=1}^{N-1}(k_N + k_i)\left(\prod_{j=1}^{N-1}\prod_{l=j+1}^{N-1}(k_j - k_l)\right)\cos\left(\theta_{\underbrace{++\cdots+\,-}_{N-1 \text{ times}}} + \alpha_{\underbrace{++\cdots+\,-}_{N-1 \text{ times}}}\right), \tag{I.18}$$

Eq. (I.14): $$\left(\prod_{i=1}^{N-1}(k_N + k_i)\prod_{i=1}^{N-2}(k_{N-1} + k_i)\left(\prod_{j=1}^{N-2}\prod_{l=j+1}^{N-2}(k_j - k_l)\right)\right)\cos\left(\theta_{\underbrace{++\cdots+\,--}_{N-2 \text{ times}}} + \alpha_{\underbrace{++\cdots+\,--}_{N-2 \text{ times}}}\right). \tag{I.19}$$

In each term, the sign within the multiplicative wave number dependent coefficients is (−) whenever $\sigma_i$ and $\sigma_j$ have the same signs, and (+) whenever they have opposite signs. This pattern repeats itself in all other terms. Hence, the general term in $\tau_E$ has the form:

$$\left(\prod_{j=1}^{N}\prod_{l=j+1}^{N}(k_j - \sigma_i\sigma_j k_l)\right)\cos(\theta_{\vec{\sigma}} + \alpha_{\vec{\sigma}}) . \tag{I.20}$$

where

$$(\vec{\sigma} = \{\sigma_1, \sigma_2, \cdots, \sigma_N\} \;,\; \sigma_1 = +1, \sigma_{i\geq 2} = \pm 1) . \tag{I.21}$$

Next, note that the same denominator appears in the definitions of all $\alpha_{\vec{\sigma}}$. Thus, in the traditional (Inverse-Scattering/Hirota) construction, the shifts in the position of soliton trajectories in the *x-t* plane have a singular dependence on the wave numbers whenever any two of them coincide.

Finally, in the traditional construction, the exponents, $\theta_i$, may include arbitrary shifts, $\delta_i$:

$$\theta_i = k_i x + \omega_i t + \delta_i , \tag{I.22}$$

The cumulative contribution of these shifts in any term,

$$\theta_{\vec{\sigma}} = \sum_{i=1}^{N} \sigma_i \theta_i , \tag{I.23}$$

is:

$$\sum_{i=1}^{N} \sigma_i \delta_i . \tag{I.24}$$

In $\tau_E$, the $2^{N-1}$ $\alpha_{\vec{\sigma}}$ constitute additional shifts in the locations of soliton trajectories in the *x-t* plane. Hence, they ought to be expressible in a similar form:

$$\alpha_{\vec{\sigma}}(\vec{k}) = \sum_{i=1}^{N} \sigma_i \alpha_i (\vec{k}) \quad , \quad (\vec{k} = \{k_1, k_2, \cdots, k_N\}) . \tag{I.25}$$

This ensures translation invariance along the trajectory of each soliton, once sufficiently far away from all other solitons. The examples of two-three and four-soliton solutions, discussed in detail in Section 2, confirm this general statements. This allows fro re-writing the equivalent $\tau$-function, $\tau_E$, in the form:

$$\tau_E(t,x) = \sum_{\vec{\sigma}} \prod_{i=1}^{N} \prod_{j=i+1}^{N} (k_j - \sigma_i \sigma_j k_i) \cosh \tilde{\theta}_{\vec{\sigma}}$$

$$\tilde{\theta}_{\vec{\sigma}} = \sum_{i=1}^{N} \sigma_i \tilde{\theta}_i \tag{I.26}$$

$$\tilde{\theta}_i = \theta_i + \alpha_i (\vec{k}) = k_i x + \omega_i t + \Delta_i (\vec{k}) \quad , \quad \Delta_i (\vec{k}) = \delta_i + \alpha_i (\vec{k})$$

The shifts, $\Delta_i(\vec{k})$, may contain the wave number dependent contributions required in the traditional case, but, as discussed in the main body of the paper, may assume any values.

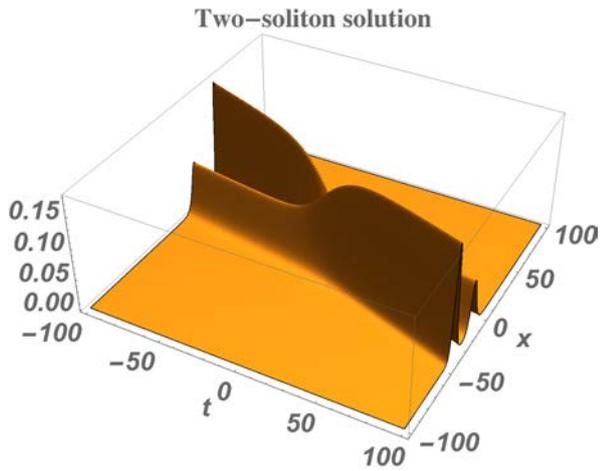

Fig. 1 Two-KdV-soliton solution (Eq. (20)). $k_1 = 0.2; k_2 = 0.3; \delta_1 = \delta_2 = \alpha = 0$.

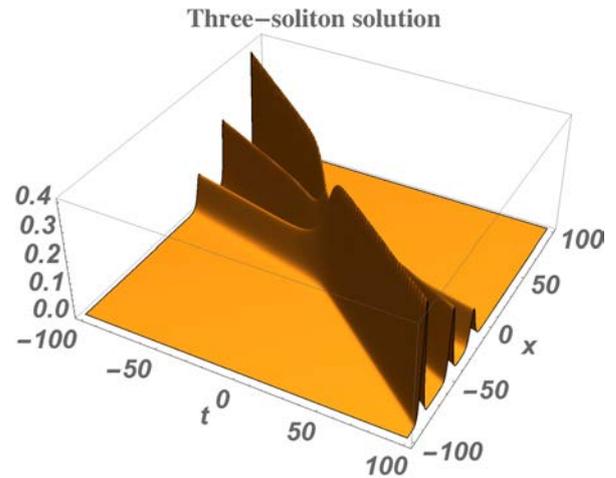

Fig. 3 Three-KdV-soliton solution (Eq. (24)). $k_1 = 0.25; k_2 = 0.35; k_3 = 0.45; \delta_1 = \delta_2 = \delta_3 = 0; \alpha_{+++} = \alpha_{++-} = \alpha_{+-+} = \alpha_{+--} = 0$.

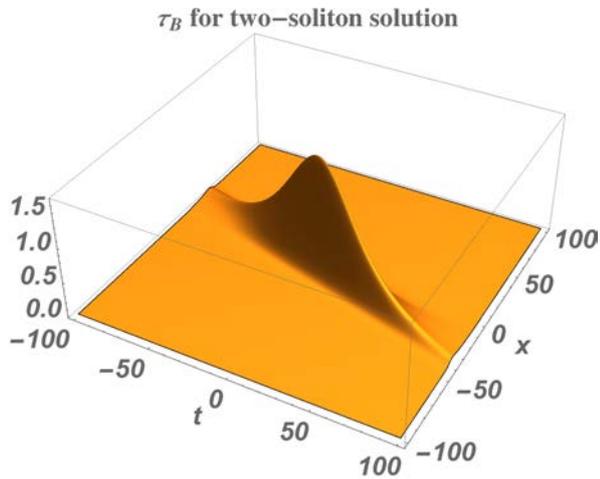

Fig. 2 Source $\tau$-function, $\tau_B$ (Eq. (21)) of two-KdV-soliton solution. Parameters as in Fig. 1.

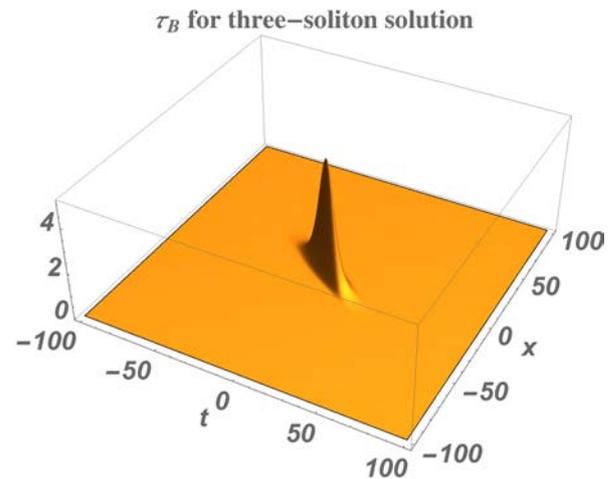

Fig. 4 Source $\tau$-function, $\tau_B$ (Eq. (34)) of three-KdV-soliton solution. Parameters as in Fig. 2.

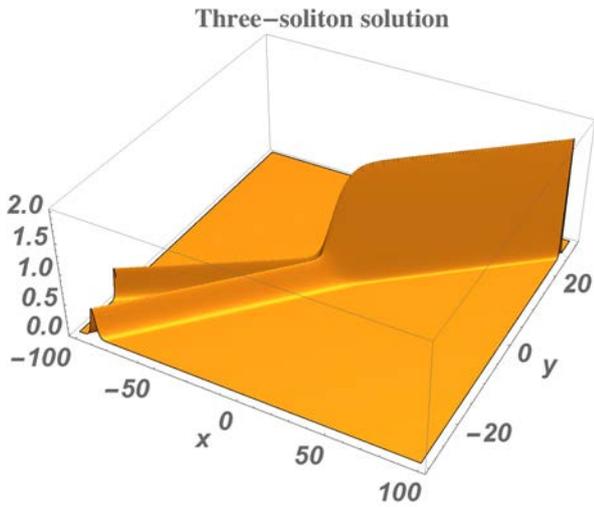
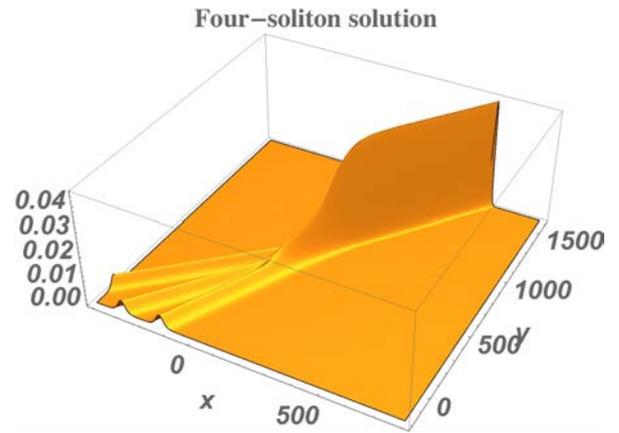

Fig. 5 (3,1) KP II-soliton solution (Eq. (58)). $k_1 = 1.; k_2 = 2.; k_3 = 3.; \xi_1 = \xi_2 = \xi_3 = 1$.

Fig.7 (4,1) KP II-soliton solution (Eq. (60)). $k_1 = 0.1; k_2 = 0.2; k_3 = 0.3; k_4 = 0.4; \xi_1 = \xi_2 = \xi_3 = \xi_4 = 1$.

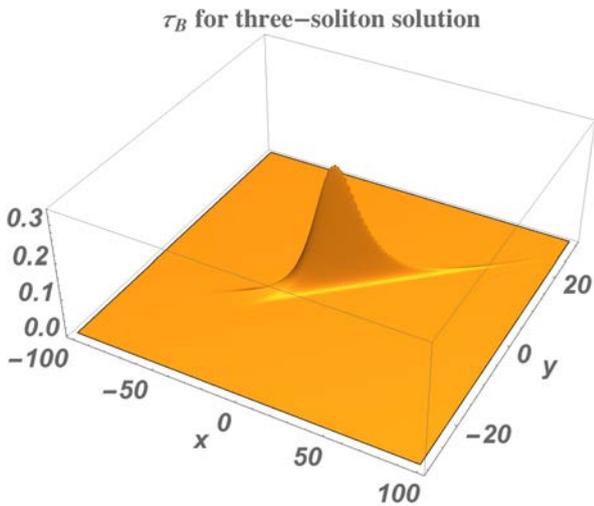
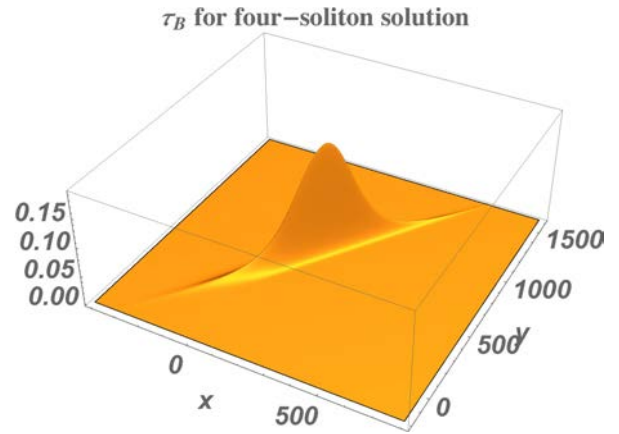

Fig. 6 Source $\tau$-function, $\tau_B$ (Eq. (48)) of (3,1)-KP II-soliton solution. Parameters as in Fig. 5.

Fig. 8 Source $\tau$-function, $\tau_B$ (Eq. (48)) of (4,1)-KP II-soliton solution. Parameters as in Fig. 7.

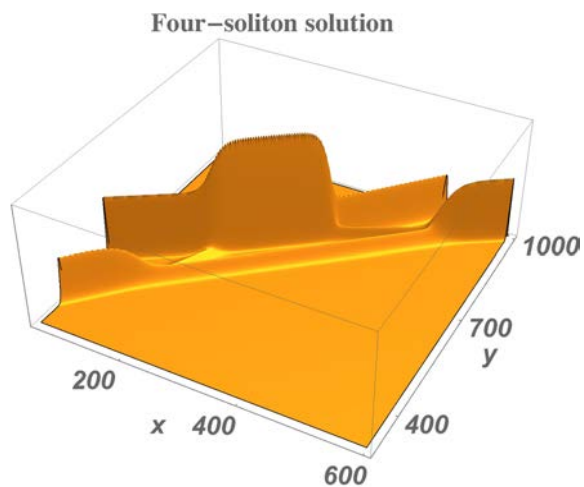

Fig. 9 (4,2) KP II-soliton solution (Eq. (62)). $k_1 = 0.1$; $k_2 = 0.3$; $k_3 = 0.6$; $k_4 = 0.9$; $\xi_{12} = 1/6$; $\xi_{13} = 4/15$; $\xi_{14} = 1/3$; $\xi_{23} = 1/10$; $\xi_{24} = 1/6$; $\xi_{34} = 1/15$; $\alpha_{++--} = \alpha_{+-+-} = \alpha_{+--+} = 0$.

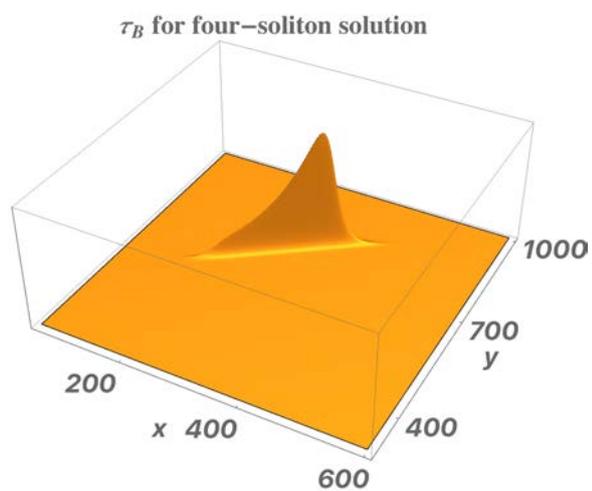

Fig. 10 Source $\tau$-function, $\tau_B$ (Eq. (48)) of (4,2)-KP II-soliton solution. Parameters as in Fig. 9.